\documentclass[onecolumn,amsmath,amssymb,11pt]{revtex4}
\usepackage{psfrag}
\usepackage{epsfig}
\usepackage{graphicx,graphics}

 \normalsize

\def\lsim{\raise0.3ex\hbox{$\;<$\kern-0.75em\raise-1.1ex\hbox{$\sim\;$}}}
\def\gsim{\raise0.3ex\hbox{$\;>$\kern-0.75em\raise-1.1ex\hbox{$\sim\;$}}}

\def\2tvec#1#2{ \left( \begin{array}{c}
#1  \\
#2  \\
\end{array} \right)}%
\def\mat2#1#2#3#4{ \left( \begin{array}{cc}
#1 & #2 \\
#3 & #4 \\
\end{array} \right) }%
\def\Mat3#1#2#3#4#5#6#7#8#9{ \left( \begin{array}{ccc}tri-bimaximal
#1 & #2 & #3 \\
#4 & #5 & #6 \\
#7 & #8 & #9 \\
\end{array} \right) } %
\def\Mat3#1#2#3#4#5#6#7#8#9{ \left(
\begin{array}{ccc}
#1 & #2 & #3 \\
#4 & #5 & #6 \\
#7 & #8 & #9 \\
\end{array} \right) }

\def\3tvec#1#2#3{ \left( \begin{array}{c}
#1  \\
#2  \\
#3  \\
\end{array} \right)}

\def\4tvec#1#2#3#4{ \left( \begin{array}{c}
#1  \\
#2  \\
#3  \\
#4  \\
\end{array} \right)}

\def\l{\lambda}%

\def\hbar{\hspace{1mm}\bar{}\hspace{-1mm}h}

  \def\m{\mu} 
   
\def\t{\theta}

\newcommand{\balg}{\begin{align}}
\def\bea{\begin{eqnarray}}
\def\eea{\end{eqnarray}} \newcommand{\be}{\begin{eqnarray}}
\newcommand{\ee}{\end{eqnarray}}
\usepackage{feynmp} \usepackage{slashed} \usepackage{latexsym}
\usepackage{graphicx} \usepackage{verbatim}
\usepackage[normalem]{ulem} 

\begin{document}

\title{Fermion masses and mixing in $\Delta(27)$ flavour model}
\author{Mohammed Abbas$^{1}$ and Shaaban Khalil$^{2,3}$} %
\vspace*{0.2cm}
\affiliation{$^1$ Department of Physics, Faculty of Science, Ain Shams University, Cairo, 11566, Egypt.\\
$^2$ Center for Fundamental Physics, Zewail City of Science and Technology, Giza 12588, Egypt.\\
$^3$Department of Mathematics, Faculty of
Science,  Ain Shams University, Cairo, 11566, Egypt.}
\date{\today}
\begin{abstract} %
\noindent
An extension of the Standard Model (SM) based on the non-Abelian discrete group $\Delta(27)$ is considered.
The $\Delta(27)$  flavour symmetry is spontaneously broken only by gauge singlet scalar fields, therefore our
 model is free from any flavour changing neural current. We show that the model accounts simultaneously for the
  observed quark and lepton masses and their mixing.
In the quark sector, we find that the up quark mass matrix is flavour diagonal and the Cabbibo-Kobayashi-Maskawa (CKM)
mixing matrix arises from down quarks .  In  the lepton sector, we show that the charged lepton mass matrix is almost diagonal.
 We also adopt type-I seesaw mechanism to generate neutrino masses.  A deviated mixing matrix from tri-bimaximal  Maki-Nakagawa-Sakata (MNS),  with $\sin\theta_{13} \sim 0.13$ and $\sin^2 \theta_{23} \sim 0.41$ , is naturally produced.

\end{abstract}
\maketitle
\section{Introduction}
The understanding of the origin of quark and lepton families and the
observed pattern of their masses and mixing is still one of the
major outstanding problems in particle physics. In the SM, these
masses and mixing are derived from Yukawa couplings, which are not
defined by the gauge symmetry. Therefore, they are arbitrary
parameters and another type of symmetry, called flavour symmetry, is
required to explain the observed fermion flavour structures. In
particular, one aims to interpret the large mass ratios between
generations: $m_u \ll m_c \ll m_t ; \quad m_d \ll m_s \ll m_b ;
\quad m_e \ll m_{\mu} \ll m_{\tau}$, and the smallness of the
off-diagonal elements of the quark weak coupling matrix, in addition
to the tiny neutrino masses and their large mixings  as recent data
suggest \cite{Capozzi:2013csa}.

Two standard approaches for dealing with flavour symmetries in
particle physics. The first one is known as "top-down" approach,
where one assumes that the SM Lagrangian is invariant under certain
flavour group $G$ and a number of Higgs-like scalar bosons, called
flavons, are coupled invariantly to SM fermions. The Vacuum
Expectation Values (VEVs) of these flavons break the flavour
symmetries and generate mass terms for SM fermions. The comparison
of the resultant mixing matrices and the mass eigenvalues with the
experimental data will confirm or refute if this group represents
the correct flavour symmetry.  In the second approach, which is
known as "bottom-up", one studies the residual symmetry that
manifests in the mass matrix and tries to relate it with the flavour
symmetry group. For instance,  by calculating the matrices $S_i$
that keep the neutrino mass matrix invariant and the matrices $T_i$
that keep the charged leptons mass matrix invariant
\cite{Hernandez:2012ra}. The group $G$ generated by these matrices
can be considered as the group of the flavour symmetry of lepton
sector.  In this regard, it was argued that for Majorana neutrinos,
regardless the form of the mass matrix $M_{\nu}$, it has $Z_2\times
Z_2$ residual symmetry \cite{Lam:2008sh,Lam:2011ag}, provided
that it has three distinct eigenvalues \cite{Grimus:2009pg}.

Many attempts were done for the interpretation of the
flavour aspects by using discrete symmetry groups (see
\cite{Hagedorn:2010th}). In particular, the non-abelian groups $A_4$
and $S_4$ have been significantly considered and shown to be useful
for obtaining tri-bimaximal neutrino mixing matrix \cite{A4, S4}.
Also $\Delta(27)$ has been considered in Ref.\cite{Ma:2007, Ma:2006,
Ross} as an example of discrete symmetries that may deviate MNS
mixing matrix from tri-bimaximal. However in \cite{Ma:2007, Ma:2006}
the attention has been devoted for lepton sector only. Also extra
Higgs (flavons) doublets have been considered, which could make the model
suffers from danger flavour changing neutral currents. It is worth noting that
within flavour symmetry approaches, the Yukawa couplings are typically generated
through non-renormalizable flavon interactions with the SM fermions,
{\it i.e.}, $Y \sim \langle \phi \rangle^n/\Lambda^n$, $n=1,2,..$.  In this respect, the  hierarchy of fermion masses
is related to the order of non-renormalizable interactions. For instance, third generation Yukawa coupling
can be obtained from $\langle \phi \rangle/\Lambda$ while first and second generation Yukawa couplings
should correspond to higher order terms.

In this  paper we explore the possibility that the flavour symmetry
based on the group $\Delta (27)$ leads to the correct quarks,
charged leptons and neutrino masses, in addition to the quark and
neutrino mixing matrices consistently with the latest experimental
results. We present a new model based on the semi-direct product
$\Delta (27)\ltimes S_2$, where $S_2$ is quite useful symmetry that
grantees the tri-bimaximal mixing as zero approximation in our
model.  We will show that deviation from tri-bimaximal neutrino mixing
matrix is related to spontaneous breaking of this symmetry.
The Higgs sector in our model consists of one Higgs doublet
only to break the electroweak symmetry and SM singlet scalars to
break the flavour symmetry. Therefore, our model is free from the
famous Flavour Changing Neural Current (FCNC) constraints that most of
constructed models suffer from, due to the existence of more than
one $SU(2)$ doublet Higgs.

We will show that the observed hierarchical structure of quark and
lepton masses can be accommodated. In addition, the small quark
mixing in the $V_{CKM}$ and large neutrino mixing in $U_{MNS}$ can
be simultaneously realised. If one assumes that left-handed quarks
and right-handed up quarks transform as triplets under $\Delta(27)$,
then one finds that the up quark mass matrix is flavour diagonal.
With right-handed downs quarks transform as singlets under
$\Delta(27)$, we will show that $V_{CKM}$ mixing matrix can be
obtained from the down quark sector.  In  the lepton sector, the
lepton doublet transform under $\Delta(27)$ as triplet, while the
right-handed charged lepton as singlets. In this case, the charged
lepton mass matrix is almost diagonal. Finally we assume
right-handed neutrinos as singlets under $\Delta(27)$, thus with the
appropriate singlet scalars (triplet and singlets under
$\Delta(27)$) we will show that a generic MNS mixing matrix can be
obtained and different interesting limits will be studied.

The paper is organised as follows.  In the next section we briefly
introduce $\Delta(27)$ flavour symmetry. In section 3 we show that
the charged lepton mass hierarchy can be naturally accounted for.
Section 4 is devoted for neutrino masses and mixing, where the
observed nearly tri-bimaximal mixing is realized.  Quark sector us
discussed in section 5. In our model the quark mixing matrix,
$V_{CKM}$, is obtained from down quarks. Finally we give our
conclusions in section 6.

\section{$\Delta(27)$ flavour symmetry}

The discrete group $\Delta (27)$ is a subgroup of $SU(3)$ and an isomorphic
to the semi-direct product group $(Z_3\times Z'_3)\ltimes Z^{\prime
\prime}_3$. It is also one of the groups $\Delta (3n^2)$ with
$n=3$. It has 27 elements and 11 conjugacy classes, so it has 11
irreducible representations, two triplets, ${\bf 3}$ and its
conjugate ${\bf \bar{3}}$, and 9 singlets ${\bf 1_1}-{\bf 1_9}$.
The group multiplication rules for $\Delta(27)$ are %
\be %
\3tvec {x_1}{x_2}{x_3}_{{\bf3}}\!\!\!\times \!\!\3tvec
{y_1}{y_2}{y_3}_{{\bf3}}\!\!\!=\!\! \3tvec {x_1 y_1}{x_2 y_2}{x_3 y_3}_{\bar
{\bf3}} \!\!\!+\!\! \3tvec {x_2 y_3}{x_3 y_1}{x_1 y_2}_{\bar {\bf3}}\!\!\!+\!\! \3tvec {x_3 y_2}{x_1 y_3}{x_2 y_1}_{\bar {\bf3}}\!\!,~
\label{multiplication1}
\ee%
and %
$3\times \bar{3}=\sum_{i=1}^9 1_i $, 
where%
\bea%
1_1&=&x_1\bar{y_1}+x_2\bar{y_2}+x_3\bar{y_3}, ~~~~~~~~~~~~~ 1_2=x_1\bar{y_1}+\omega
x_2\bar{y_2}+\omega^2x_3\bar{y_3}, \nonumber\\
1_3 &=&x_1\bar{y_1}+\omega^2x_2\bar{y_2}+\omega x_3\bar{y_3},~~~~~~~~
1_4=x_1\bar{y_2}+x_2\bar{y_3}+x_3\bar{y_1}, \nonumber\\
1_5&=&x_1\bar{y_2}+\omega x_2\bar{y_3}+\omega^2
x_3\bar{y_1},  ~~~~~~~~ 1_6=x_1\bar{y_2}+\omega^2 x_2\bar{y_3}+\omega x_3\bar{y_1},\nonumber\\
1_7&=&x_2\bar{y_1}+ x_3\bar{y_2}+ x_1\bar{y_3}, ~~~~~~~~~~~~~
1_8=x_2\bar{y_1}+\omega^2 x_3\bar{y_2}+\omega x_1\bar{y_3},\nonumber\\
1_9&=&x_2\bar{y_1}+\omega x_3\bar{y_2}+\omega^2 x_1\bar{y_3},
 \label{multiplication2}\eea
where $\omega =e^{2\pi i/3}$. The singlets multiplications are given
in Table \ref{table1}

\begin{table}[t]
\centering {\fontsize{10}{12}
\begin{tabular}{|c|c|c|c|c|c|c|c|c|c|} \hline
           &~ ${\bf 1_2}$ ~ & ~ ${\bf 1_3}$ ~ & ~ ${\bf 1_4}$ ~
              & ~ ${\bf 1_5}$ ~ &~ ${\bf 1_6}$~ & ~ ${\bf 1_7}$ ~ & ~ ${\bf 1_8}$ ~ & ~ ${\bf 1_9}$~ \\ \hline
~${\bf 1_2}$ ~    &${\bf 1_3}$&      &        &
              &     &    &         &     \\ \hline
${\bf 1_3}$  &${\bf 1_1}$& ${\bf 1_2}$      &        &
              &     &     &        &     \\ \hline
${\bf 1_4}$     &${\bf 1_6}$&    ${\bf 1_5}$      &${\bf 1_7}$   &
              &    &  &   &    \\ \hline
${\bf 1_5}$    &${\bf 1_4}$& ${\bf 1_6}$      &${\bf 1_9}$ &${\bf
1_8}$
              &     &   &   &     \\ \hline
${\bf 1_6}$   &${\bf 1_5}$   &${\bf 1_4}$      &${\bf 1_8}$   &${\bf
1_7}$
              &${\bf 1_9}$      &   &  &     \\ \hline
${\bf 1_7}$&  ${\bf 1_8}$&  ${\bf 1_9}$      &${\bf 1_1}$ &${\bf
1_3}$
              &${\bf 1_2}$      &${\bf 1_4}$ &     &  \\ \hline
${\bf 1_8}$&  ${\bf 1_9}$&  ${\bf 1_7}$      &${\bf 1_2}$ &${\bf
1_1}$
              &${\bf 1_3}$      &${\bf 1_6}$ & ${\bf 1_5}$    &  \\ \hline
${\bf 1_9}$&  ${\bf 1_7}$&  ${\bf 1_8}$      &${\bf 1_3}$ &${\bf
1_2}$
              &${\bf 1_1}$      &${\bf 1_5}$ & ${\bf 1_4}$    & ${\bf 1_6}$ \\ \hline
\end{tabular}}
\caption{The singlet multiplications of the group $\Delta(27)$}.
\label{table1}
\end{table}

Non-vanishing neutrino masses imply the existence of three right
handed neutrinos. Therefore, we consider the matter sector of SM
besides three  right handed neutrinos. We assign the lepton doublet
to the triplet ${\bf 3}$ of $\Delta(27)$, while right handed
components are ascribed to different singlet representations of
$\Delta (27)$.  As mentioned, we consider only one SM Higgs scalar
$(H)$ and the following singlets: $\chi$, $\xi$, $\eta$, $\sigma$
and $\phi$ that break the flavour symmetry.

In order to get the tri-bimaximal mixing as zero order approximation in our model, we find that
an additional $S_2$ symmetry should be considered. The $S_2$, the group of permutation of two objects, has the
following generators in the 3-dimensional representation:
\be%
 e=\left(
      \begin{array}{ccc}
        1 & 0 & 0 \\
        0 & 1 & 0 \\
        0 & 0 & 1 \\
      \end{array}
    \right)~~~~~~a=\left(
                                       \begin{array}{ccc}
                                         1 & 0 & 0 \\
                                         0 & 0 & 1 \\
                                         0 & 1 & 0 \\
                                       \end{array}
                                     \right).
\ee
The particle transformations under $S_2$ are given by:
\be%
f_1 \leftrightarrow f_1,~~~~~~~~f_2 \leftrightarrow f_3,
\ee%
where $f$ stands for $L_i$, $\chi_i$, $\eta_i$, $\phi$, $\xi_i$,
$\sigma$ and $\nu_{R_i}$. The right handed charged leptons transform
trivially under $S_2$ symmetry. Moreover, we consider extra $Z_4$
group to get the correct mass hierarchy of the charged leptons. In
Table \ref{table1}, we present the field transformation under
$\Delta(27)$ and $Z_4$.

Before concluding this section, we comment on possible vacuum alignments for the VEVs of the singlet scalars.
The $\Delta (27)\ltimes S_2$ invariant scalar potential at the renormalizable level is given by:
\bea%
V&=&m_1^2 \eta_1^{\dagger}\eta_1+
g_1(\eta_1^{\dagger}\eta_1)^2+m_2^2(\eta_2^{\dagger}\eta_3+\eta_3^{\dagger}\eta_2)+g_2\eta_1^{\dagger}\eta_1
\eta_2^{\dagger}\eta_3+g_3\eta_2^{\dagger}\eta_2\eta_2^{\dagger}\eta_1+g_4\eta_3^{\dagger}\eta_3\eta_3^{\dagger}\eta_1
\nonumber\\&+&g_5\eta_2^{\dagger}\eta_2\eta_3^{\dagger}\eta_3+m_{\xi}^2\xi^{\dagger}\xi
+h_1(\xi^{\dagger}\xi)^2+m_{\sigma}^2\sigma^{\dagger}\sigma+h_2(\sigma^{\dagger}\sigma)^2+m_{\phi}^2\phi^{\dagger}\phi+
h_3(\phi^{\dagger}\phi)^2\nonumber\\&+&m_{\chi}^2\chi^{\dagger}\chi+
h_4(\chi^{\dagger}\chi)^2+h_5\eta_1^{\dagger}\eta_1\xi^{\dagger}\xi+h_6\eta_1^{\dagger}\eta_1\sigma^{\dagger}\sigma+
h_7\eta_1^{\dagger}\eta_1\phi^{\dagger}\phi
+h_8\eta_1^{\dagger}\eta_1\chi^{\dagger}\chi\nonumber\\&+&h_9\eta_2^{\dagger}\eta_3\xi^{\dagger}\xi+h_{10}\eta_2^{\dagger}\eta_3\sigma^{\dagger}\sigma+
h_{11}\eta_2^{\dagger}\eta_3\phi^{\dagger}\phi
+h_{12}\eta_2^{\dagger}\eta_3\chi^{\dagger}\chi+h_{13}\xi^{\dagger}\xi\phi^{\dagger}\phi+h_{14}\xi^{\dagger}\xi\chi^{\dagger}\chi
\nonumber\\&+&h_{15}\phi^{\dagger}\phi\chi^{\dagger}\chi+h_{16}\xi^{\dagger}\xi\sigma^{\dagger}\sigma+h_{17}\chi^{\dagger}\chi\sigma^{\dagger}\sigma+
h_{18}\phi^{\dagger}\phi\sigma^{\dagger}\sigma +h.c.
\eea
\begin{table} \begin{tabular}{|c|c|c|c|c|c|c|c|c|c|c|c|c|c|}
  \hline
  ~ Fields ~ & ~ $\ell$ ~ & ~ $e_R$ ~ &$\mu_R$ ~ & ~ $\tau_R$ ~ & ~ $\nu_{R_\alpha}$ ~ & ~ $H$ ~ & ~ $\chi$ ~ & ~ $\eta_\alpha$~ &~ $\xi$~ &~ $\phi$~ &~ $\sigma$ ~  \\
  \hline $\Delta (27)$&3&$1_1$ &$1_1$&$1_1$& $1_\alpha$ & $1_1$ & $3$&$1_\alpha$&$\bar{3}$&3 &$\bar{3}$  \\
  \hline $Z_4$ & 1 &1 &-1&$i$& $-i$ & 1 & $i$ & -1 & $i$ &$-i$&-1 \\
  \hline
\end{tabular}
\centering
\caption{Field transformations under
$\Delta (27)$, and $Z_4$.  Here $\alpha$ refers to $1, 2, 3$.}
\label{table2}
\end{table}
The $S_2$ symmetry leads to $g_3=g_4$.  From this equation, one can notice that the potential contains 6 free mass parameters and 23 free self-interacting couplings. This large number of free parameters is one of the features of any flavour scalar (non-supersymmetric) potential. Therefore, the minimization conditions
of this potential can imply the following  extremum solutions:
\bea%
\langle\xi\rangle&=&(0, w, 0), ~~~~~\langle\phi\rangle=(0, 0,
w^\prime),~~~~~\langle\chi\rangle=(v, v, v)\nonumber\\
\langle\sigma\rangle &=&(v^\prime, v^\prime, v^\prime),
~~~~~~\langle\eta_1\rangle=u_1, ~~~\langle\eta_2\rangle=
u_2,~~~\langle\eta_1\rangle= u_3\label{vevs alignments chl1}
\eea%
with the conditions: \bea u_1^2 & \sim &u_2^2\sim u_3^2\sim w^2\sim
w^{\prime^2}\sim v^{\prime^2} \sim v^2 \nonumber\\ &\sim &
-\frac{m_1^2}{k_1}\sim -\frac{m_2^2}{k_2}\sim -\frac{m_2^2}{k_3}\sim
-\frac{m_{\xi}^2}{k_4}\sim -\frac{m_{\phi}^2}{k_5}\sim
-\frac{m_{\chi}^2}{k_6}\sim -\frac{m_{\sigma}^2}{k_7} , \eea where
\bea
k_1&=&2g_1+g_2+2g_3+h_5+3h_6+h_7+3h_8, \nonumber\\
k_2&=&g_2^*+g_3+g_5+h_9^*+3h_{10}^*+h_{11}^*+3h_{12}^*, \nonumber\\
k_3&=&g_2+g_3+g_5+h_9+3h_{10}+h_{11}+3h_{12}, \\
k_4&=&2h_1+h_{13}+3h_{14}+h_5+h_9+3h_{16}, \nonumber\\
k_5&=&2h_3+h_{13}+3h_{15}+h_7+h_{11}+3h_{18}, \nonumber\\
k_6&=&8h_4+h_{14}+h_{15}+h_8+h_{12}+3h_{17}, \nonumber\\
k_7&=&8h_2+h_{18}+h_{16}+h_6+h_{10}+3h_{17}, \nonumber \eea

\section{charged Lepton masses and $Z_4$ symmetry}
As shown in Table \ref{table2}, the lepton doublet is assigned to
the triplet ${\bf 3}$ of $\Delta(27)$ while right-handed lepton
$l^c_i$ are ascribed as singlet representations of $\Delta (27)$.
We find that the hierarchy between the charged lepton masses may be achieved by
imposing an extra discrete symmetry $Z_4$.  A possible set of charge
assignments of $Z_4$ that lead to Yukawa interactions compatible
with
the experimental data is also given in Table \ref{table2}. Therefore, the charged lepton Yukawa Lagrangian, invariant under $\Delta (27)\ltimes S_2 \times Z_4$ is given by%
\bea%
\!{\cal L}_l\!=\!\frac{\lambda_e}{\Lambda^4} \bar{\ell} H
e_{R}(\chi^4\!+\!\phi^4)\!\!+\!\! \frac{\lambda_{\mu}}{\Lambda^2} \bar{\ell} H
\mu_{R} (\xi\xi)\!\!+\!\! \frac{\lambda_{\tau}}{\Lambda} \bar{\ell} H
\tau_{R} (\phi)
\!\!+\!\! h.c.,%
\label{lag.1}\eea
where $\Lambda$ is non-renormalization scale, which is  much larger than TeV.
The Yukawa couplings $\lambda_{e}, \lambda_{\mu}, \lambda_{\tau}$
are of order one. As mentioned in the previous section, the scalar
potential $V(\phi,\chi,\xi)$ contains several free parameters that
can be adjusted to generate the VEVs for the flavons as given in
Eq.(\ref{vevs alignments chl1}).  To get the hierarchal mass spectra
of charged leptons, one assumes that the VEVs $w, w^\prime,
v^\prime$ and $v$ are of the same order and satisfy the following
relation:
\bea%
\frac{w}{\Lambda}\sim
\frac{w^\prime}{\Lambda}\sim\frac{v}{\Lambda}\sim\frac{v^\prime}{\Lambda}\sim
{\cal
O}(\lambda_C^2),\label{scaling} \eea%
where $\lambda_C$ is the Cabibbo angle, {\it i.e},  $\lambda_C \sim 0.22$.
In this case, one finds that
the charged lepton mass matrix $m_{\ell}$ is given by
\bea%
m_{\ell}=\left(
  \begin{array}{ccc}
    \lambda_C^6 & 0 & 0 \\
    \lambda_C^6 & \lambda_C^2 & 0 \\
    \lambda_C^6 & 0 & 1 \\
  \end{array}
\right)\lambda_\tau \lambda_C^2\langle H\rangle.
\eea%
The matrix $m_{\ell}$ is not symmetric or Hermition, so it can be
diagonalized by two unitary matrices. %
\be%
m_{\ell} = U_L^T~ m_{\ell}^{diag} U_R =\lambda_{\tau}\lambda^2_C
\langle H\rangle\!\!\left(
     \begin{array}{ccc}
       1 \!&\! -\lambda_C^8 \!&\! 0 \\
       \lambda_C^8 \!&\! 1 \!&\! 0 \\
       0 \!&\! 0 \!&\! 1 \\
     \end{array}
   \!\right)\!\left(\!
            \begin{array}{ccc}
              \lambda_C^6 \!&\! 0 \!&\! 0 \\
              0 \!&\! \lambda_C^2 \!&\! 0 \\
              0 \!&\! 0 \!&\! 1 \\
            \end{array}
          \!\right)\!\left(\!
           \begin{array}{ccc}
             1 \!&\! \lambda_C^4 \!&\! \lambda_C^6 \\
             -\lambda_C^4 \!&\! 1 \!&\! 0 \\
             -\lambda_C^6 \!&\! 0 \!&\! 1 \\
           \end{array}
         \!\right)\!.~~~~
         \ee
Therefore the charged lepton masses are given by:
\bea
m_e&\sim &\lambda_\tau\lambda_C^8\langle H \rangle, \nonumber\\
m_{\mu}&\sim&\lambda_\tau\lambda_C^4\langle H \rangle , \\
m_{\tau}&\sim&\lambda_\tau\lambda_C^2\langle H \rangle\nonumber.
\eea Hence, the following mass relations  are satisfied
\bea%
m_\tau&:& m_\mu~ :~ m_e ~\approx~ 1~: ~\lambda_C^2 ~:~ \lambda_C^6,
\eea%
which are consistent with the hierarchy between the charged lepton masses:
$$m_e = 0.511~{\rm MeV} , ~~~~~~  m_{\mu} = 105.658~ {\rm MeV}, ~~~~~~
m_{\tau} = 1.776~ {\rm GeV}.$$
It is worth noting that the left handed mixing matrix $U_L$ is close to the identity
matrix, so the lepton mixing should arise mainly from the neutrino sector.

\section{Neutrino masses and mixing}
From solar and atmospheric neutrino oscillation data
\cite{Girardi:2013zra},  the neutrino mass squared differences are
given by:
\be
\Delta m^2_{21} = 7.54^{+0.26}_{-0.22} \times 10^{-5}
{\rm eV^2}, ~~  \vert \Delta m^2_{31}\vert =
2.47^{+0.06}_{-0.22} \times 10^{-3} {\rm eV^2}, ~~  \vert \Delta
m^2_{32} \vert = 2.46^{+0.07}_{-0.11} \times 10^{-3} {\rm eV^2}.~~ \ee
In addition, the latest best-fit results for the mixing pattern in
the lepton sector is given by \cite{Capozzi:2013csa}
\be%
\sin^2 \theta_{12} = 0.308^{+0.017}_{-0.017}, ~~~~ \sin^2
\theta_{23} = 0.437^{+0.033}_{-0.023}, ~~~~ \sin^2 \theta_{13} =
0.0234^{+0.0020}_{-0.0019}.%
\ee%

Having the lepton doublet, $\ell_i$, as $\Delta(27)$ triplet and
right-handed neutrinos, $\nu_{R_j}$, as singlets, then one can
construct the following invariant interaction terms: \be {\cal L}_D=
\frac{1}{\Lambda} \lambda_{ijk} \bar{\ell}_i \nu_{R_j} H \chi_k,
\ee%
where $\chi$ is $\Delta(27)$ triplet scalar.  Therefore, one gets the following terms:
\bea
&& \frac{1}{\Lambda}\lambda_1 (\bar\ell_1 \chi_1+\bar\ell_2
\chi_2+\bar\ell_3 \chi_3) \nu_{R_1}H,\nonumber\\
&& \frac{1}{\Lambda}\lambda_2 (\bar\ell_1 \chi_1+\omega^2 \bar\ell_2 \chi_2+\omega \bar\ell_3
\chi_3) \nu_{R_2}H,\nonumber\\
&& \frac{1}{\Lambda}\lambda_3 (\bar\ell_1 \chi_1+\omega \bar\ell_2 \chi_2+\omega^2 \bar\ell_3 \chi_3)
\nu_{R_3}H. \nonumber
\eea
The $S_2$ flavour symmetry imposes the equality
of the second and third couplings: $\lambda_2=\lambda_3$.
After the flavour symmetry breaking through the aligned vacuum:
$\langle\chi\rangle=(v, v, v)$, the following Dirac neutrino mass matrix is obtained
\bea
m_D=\frac{v}{\Lambda}\left(
  \begin{array}{ccc}
    \lambda_1  & \lambda_2  & \lambda_2  \\
    \lambda_1  & \omega^2 \lambda_2  &\omega \lambda_2  \\
    \lambda_1  & \omega \lambda_2  & \omega^2 \lambda_2  \\
  \end{array}
\right)\langle H \rangle, \label{dirac matix}
\eea
which can be expressed as
\bea
m_D=
\frac{v}{\Lambda} \left(
                 \begin{array}{ccc}
                   1 & 1 & 1 \\
                   1 & \omega^2 & \omega \\
                   1 & \omega & \omega^2 \\
                 \end{array}
               \right) \left(
                        \begin{array}{ccc}
                          \lambda_1 & 0 & 0 \\
                          0 & \lambda_2 & 0 \\
                          0 & 0 & \lambda_2 \\
                        \end{array}
                      \right)\langle H \rangle .\label{dirac magic}
\eea %
Note that here all Dirac neutrino masses are generated from the same non-renormalizable interactions of order $v/\Lambda$. Therefore, one would not expect any hierarchy between them.

Furthermore from the invariant interactions of right-handed neutrinos with $\Delta(27)$ singlets $\eta_i$, Majorana mass terms for $\nu_R$ can be obtained from the following renormalizable interactions:
\bea%
{\cal L}_M=f_{ijk}\bar{\nu}^c_{R_i} \nu_{R_j}\eta_k.
\eea
According to the $\Delta(27)$ multiplication rules of singlet representations, the invariants that
give right-handed neutrino masses are:
\bea
&& f_1\bar{\nu}^c_{R_1}
\nu_{R_1}\eta_1,~~~~~~ f_2\bar{\nu}^c_{R_1}
\nu_{R_2}\eta_3,~~~~~~
f_3\bar{\nu}^c_{R_1} \nu_{R_3}\eta_2,\nonumber\\
&&f_4\bar{\nu}^c_{R_2} \nu_{R_2}\eta_2,~~~~~~
f_5\bar{\nu}^c_{R_2} \nu_{R_3}\eta_1,~~~~~ f_6\bar{\nu}^c_{R_3}
\nu_{R_3}\eta_3.
\eea
The symmetry $S_2$  imposed the following constraints:
$$f_2=f_3~~~~~~~~~~~~~~~~~~~f_4=f_6.$$
Therefore, after $\Delta(27)$ symmetry breaking through the VEVs of $\eta_k$, one obtains the following right-handed Majorana mass
matrix:
\bea
M_R=\left(
  \begin{array}{ccc}
    f_1 u_1 & f_3 u_3 & f_3 u_2 \\
    f_3 u_3 & f_4 u_2 & f_5 u_1 \\
    f_3 u_2 & f_5 u_1 & f_4 u_3 \\
  \end{array}
\right).\label{mR} \eea %

As usual, the light neutrino mass matrix is obtained in terms of Dirac neutrino mass matrix and right-handed neutrino through type I seesaw mechanism as
\bea
M_{\nu}=- m_D M_R^{-1} m_D^T.
\label{mn}
\eea
It is noticeable  that the mass matrix $M_R$  in Eq.(\ref{mR}) is generic matrix that can lead to different type of neutrino mixing matrix (tri-bimaximal or nearly  tri-bimaximal mixing matrix), depending on the coupling $f_3$ and the difference between the VEVs $u_2$ and $u_3$. In general the tri-bimaximal mixing matrix, $U_{TBM}$, can be
written as \cite{Ma:Uw, Ma:2004}:
\bea%
U_{TBM}=\Gamma_{mag} U^{\prime},\label{u_tbm} \eea where
$\Gamma_{mag}$ is the magic matrix proposed by Cabibbo
\cite{cabibo:1978} and Wolfenstein \cite{wolfenstein} and has the
form:
\bea%
\Gamma_{mag}=\frac{1}{\sqrt{3}}\left(
                 \begin{array}{ccc}
                   1 & 1 & 1 \\
                   1 & \omega & \omega^2 \\
                   1 & \omega^2 & \omega \\
                 \end{array}
               \right),\label{magec matrix}
               \eea%
 and
               \bea%
                U^{\prime}=\left(
                                                  \begin{array}{ccc}
                                                    1 & 0 & 0 \\
                                                    0 & \frac{1}{\sqrt{2}} & \frac{1}{\sqrt{2}} \\
                                                    0 & \frac{1}{\sqrt{2}} & -\frac{1}{\sqrt{2}} \\
                                                  \end{array}
                                                \right)\left(
                                                         \begin{array}{ccc}
                                                           0 & 1 & 0 \\
                                                           1 & 0 & 0 \\
                                                           0 & 0 & i \\
                                                         \end{array}
                                                       \right)=\left(
                                                                 \begin{array}{ccc}
                                                                   0 & 1 & 0 \\
                                                                   \frac{1}{\sqrt{2}} & 0 & \frac{i}{\sqrt{2}} \\
                                                                   \frac{1}{\sqrt{2}} & 0 & -\frac{i}{\sqrt{2}} \\
                                                                 \end{array}
                                                               \right).
                                                \eea
From Eqs. (\ref{dirac
magic}) and (\ref{magec matrix}):
\bea%
M_{\nu}=-\frac{3 v^2}{\Lambda^2} \langle H
\rangle^2 \Gamma_{mag} D_{\lambda}M_R^{-1}D_{\lambda}\Gamma_{mag},
\label{mnu}
\eea%
where $D_{\lambda}=diag(\lambda_1, \lambda_2, \lambda_2)$.
If $M_\nu$ is diagonalized by tri-bimaximal mixing matrix, then we can determine
the corresponding form of the right-handed neutrino mass matrix, which typically takes the form:
\bea%
(M_R)_{TBM}=\left(
  \begin{array}{ccc}
    x & 0 & 0 \\
    0 & z & y \\
    0 & y & z \\
  \end{array}
\right).\label{right mass}
\eea
Therefore, the exact tri-bimaximal can be naturally obtained within $\Delta(27)$ flavour symmetry if the coupling $f_3=0$ and the VEVs $u_2=u_3=u$, which ensures the $S_2$ invariance.  In this case, one obtains
\bea
M_\nu^{\mbox{diag}}=-3 \frac{v^2}{\Lambda^2} \langle H \rangle^2\left(
\begin{array}{lll}
 \frac{ \lambda_2^2 }{{f_5} {u_1}+{f_4} {u}} & 0 & 0 \\
 0 & \frac{ \lambda_1^2 }{{f_1} {u_1}} & 0 \\
 0 & 0 & \frac{ \lambda_2^2 }{{f_5} {u_1}-{f_4} {u}}
\end{array} \right).
\label{TBM_masses}
\eea
As expected, unlike the charged lepton masses, here there is no clear argument for neutrino mass hierarchy. Instead one
should assume a hierarchy among the involved couplings of flavon VEVs to achieve the type of desired neutrino mass spectrum.
For instance if one considers $f_4\sim f_5\gg f_1$, $u_1\sim u$
and the couplings $\lambda_s$ are of the same order, the normal neutrino mass
hierarchy is realized. While inverted neutrino mass hierarchy is obtained if
$f_4\sim f_5\gg f_1$ and $u_1\sim -u$. Finally, degenerate scenario
is obtained if $f_1\sim f_5\gg f_4$ and $u_1\gg u$.

Now we consider the case of $f_3\neq 0$ and  $u_2=u_3=u$ ({\it i.e.}, $M_\nu$ is still invariant under $S_2$ symmetry).  In this case, the neutrino mass matrix is given by%
\bea%
M_\nu=\frac{v^2}{\Lambda^2} \langle H \rangle^2\left(
              \begin{array}{ccc}
                A & B & B \\
                B & C & D \\
                B & D & C \\
              \end{array}
            \right),
            \eea
where,
\bea%
A&=&\frac{f_5 u_1 \l_1^2+(f_4 \l_1-4 f_3 \l_2) u \l_1+2 f_1 \l_2^2
u_1}{f_1 u_1 (f_5 u_1+f_4 u)-2 f_3^2 u^2},\nonumber\\
B&=&\frac{f_5 u_1 \l_1^2+(f_4 \l_1- f_3 \l_2) u \l_1- f_1 \l_2^2
u_1}{f_1 u_1 (f_5 u_1+f_4 u)-2 f_3^2 u^2},\nonumber\\
C&=&\frac{f_5^2 \l_1^2 u_1^2 + 2 f_5 \l_2 u_1 (f_1 \l_2 u_1 + f_3
\l_1 u) - u (-f_1 f_4 \l_2^2 u_1 + (f_4^2 \l_1^2 + 2 f_3 f_4 \l_1
\l_2 + 3 f_3^2 \l_2^2) u)}{(f_5 u_1 - f_4 u)(f_1 u_1 (f_5 u_1+f_4
u)-2
f_3^2 u^2)},\nonumber\\
D&=&\frac{f_5^2 \l_1^2 u_1^2 +  f_5 \l_2 u_1 (-f_1 \l_2 u_1 +2 f_3
\l_1 u) - u (2f_1 f_4 \l_2^2 u_1 + (f_4^2 \l_1^2 + 2 f_3 f_4 \l_1
\l_2 - 3 f_3^2 \l_2^2) u)}{(f_5 u_1 - f_4 u)(f_1 u_1 (f_5 u_1+f_4
u)-2
f_3^2 u^2)}.\nonumber
\eea%
As emphasised in Ref.\cite{Abbas:2010jw}, the tri-bimaximal mixing matrix corresponds to neutrino mass matrix that satisfies the following three
conditions:%
\bea%
(M_\nu)_{12}&=&(M_\nu)_{13},\nonumber\\
(M_\nu)_{22}&=&(M_\nu)_{33}, \nonumber\\
(M_\nu)_{11}+(M_\nu)_{12}&=&(M_\nu)_{22}+(M_\nu)_{23}
\label{two conditions}
\eea%
It is clear that the neutrino mass matrix in our case
satisfies the first two conditions only while the third condition is violated. 
Therefore, this neutrino mass matrix can be
diagonalized by a matrix which is very close to tri-bimaximal. However, we find that the resulting mixing matrix still has zero $\t_{13}$ and maximal $\t_{23}$. It essentially deviates from tri-bimaximal in the first and the second columns only. Also the corresponding  eigenvalues of neutrino masses are given by%
\bea%
m_1&=&-3 \frac{v^2}{\Lambda^2} \langle H \rangle^2\frac
{2\lambda_1^2\l_2^2}{(
 x +
  \sqrt{x^2 -
   y^2})},\nonumber\\m_2&=&-3 \frac{v^2}{\Lambda^2} \langle H \rangle^2\frac {2\lambda_1^2\l_2^2}{(
x -
  \sqrt{x^2 -
   y^2})},\nonumber\\m_3&=&-3 \frac{v^2}{\Lambda^2} \langle H \rangle^2\frac{\lambda_2^2}{(f_5 u_1-f_4 u )}.
   \eea%
   where
     $x=(f_1\l_2^2  + f_5\l_1^2) u_1 + f_4 u\l_1^2$ and $y^2=4\l_1^2\l_2^2 (-2 f_3^2 u^2 + f_1 u_1 (f_5 u_1 + f_4
   u))$.  Here the normal hierarchy is achieved if $f_4\sim f_5\gg f_1\gg f_3$, $u_1\sim u$ and the couplings $\lambda_1\sim \lambda_2$. The degenerate scenario is obtained if $f_1\sim f_5\gg f_4\gg f_3$ and $u_1\gg u$. Finally, the inverted hierarchy is obtained if $f_4\sim f_5\gg f_1\gg f_3$ and $u_1\sim -u$.

Now we turn to the case of spontaneous $S_2$ symmetry breaking, {\it i.e}, $u_2\neq u_3$ with $f_3\sim 0$.  In this case, all the three relations, in Eq.(\ref{two conditions}), are violated. The consequences of the
deviation from tri-bimaximal  mixing on the symmetry manifesting in
neutrino mass matrix were studied in \cite{Abbas:2010jw}. Following
the notations used in this reference, we define the parameters which
characterize the deviation of mixing angles from
the tri-bimaximal  values as %
\bea %
D_{12}\equiv \frac{1}{3}-s_{12}^2,~~~~~~
D_{23} \equiv \frac{1}{2}-s_{23}^2, ~~~~~~ D_{13}\equiv s_{13},
\label{TBMdeviation} %
\eea%
where $s_{ij}\equiv\sin\theta_{ij}$. The violation of tri-bimaximal  symmetry
of neutrino mass matrix in Eq. (\ref{two conditions}) can be written
in terms of deviation parameters $D_{23}$ and $s_{13}$ as follows: %
\begin{widetext}
\bea%
\Delta_1&=&(M_\nu)_{12}-(M_\nu)_{13}=\frac{\sqrt{2}}{3}((2 m_1+m_2)e^{2i\delta}-3m_3 )s_{13}e^{-i\delta}+\frac{2}{3}(m_2-m_1)D_{23},\nonumber\\%
\Delta_2&=&(M_\nu)_{22}-(M_\nu)_{33}=\frac{2\sqrt{2}}{3}(m_2-m_1)s_{13}e^{i\delta}+\frac{1}{3}(m_1+2m_2-3m_3)D_{23},\nonumber\\
   \Delta_3&=&(M_\nu)_{11}+(M_\nu)_{12}-((M_\nu)_{22}+(M_\nu)_{23})\nonumber\\
   &=& \Big(\frac{1}{3\sqrt{2}}(3m_3-(2m_1+m_2)e^{2i\delta})e^{-i\delta}-
   \frac{\sqrt{2}}{3}(m_2-m_1)e^{i\delta}\Big)
   s_{13}\nonumber\\&+&\Big(\frac{2}{3}(3m_3-(2m_1+m_2)e^{2i\delta})e^{-2i\delta}-\frac{1}{3}(m_2-m_1)\Big)
   \frac{s^2_{13}}{2}\nonumber\\&-&\frac{1}{3}(2m_1+m_2-3m_3)D_{23}-\frac{9}{4}(m_2-m_1)D_{12},\label{Delta1}
   \eea%
   \end{widetext}
where $m_i$ are the masses of effective neutrino and $\delta$ is the
leptonic Dirac phase. In our model the deviations from tri-bimaximal
conditions in Eq. (\ref{two conditions}) can give constrains on our
parameters (couplings and VEVs) in order to get
the correct mixing angles and desired scenario of mass spectra,
\bea%
\Delta_1&=&-\frac{v^2}{\Lambda^2} \langle H \rangle^2\frac{i
\sqrt{3} \l_2^2 (u_2-u_3) f_4 }{f_5^2 u_1^2-f_4^2 u_2
   u_3},\nonumber\\
\Delta_2&=&\frac{v^2}{\Lambda^2} \langle H \rangle^2\frac{i
\sqrt{3}\l_2^2 (u_2-u_3) f_4 }{ f_5^2 u_1^2-f_4^2 u_2
   u_3},\nonumber\\
   \Delta_3&=&-\frac{v^2}{\Lambda^2} \langle H \rangle^2\frac{i \sqrt{3}
\l_2^2 (u_2-u_3) f_4 }{f_5^2 u_1^2-f_4^2 u_2
   u_3}. \label{Delta2}\eea%

\begin{figure}[t]
\epsfig{file=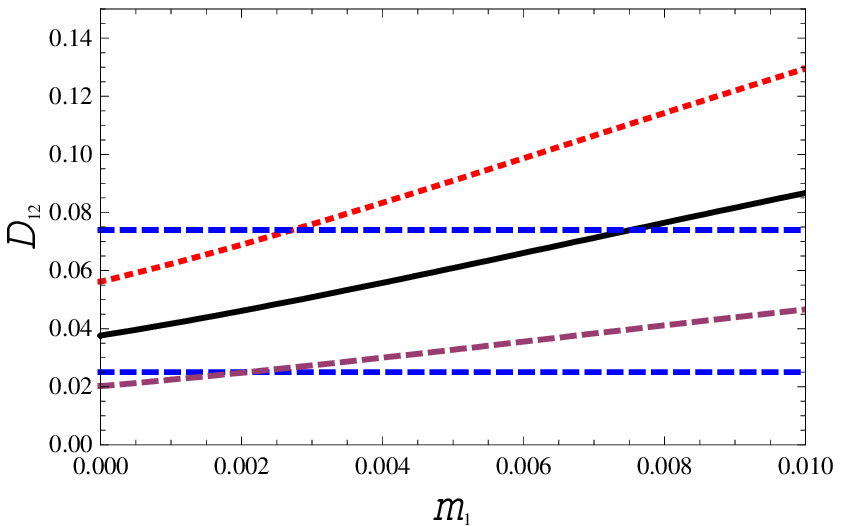,height=6.0cm,width=8.0cm,angle=0}~~~\epsfig{file=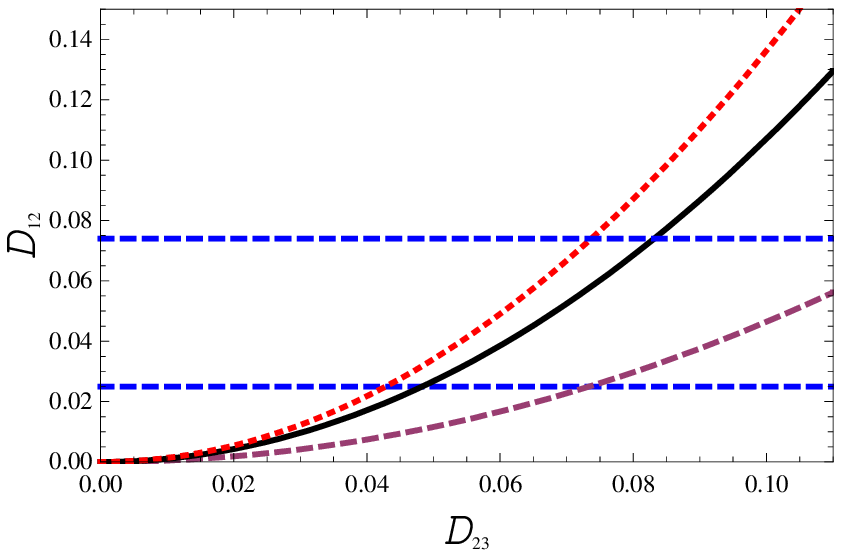,height=6.0cm,width=8.0cm,angle=0}
\caption{(Left) The deviation parameter $D_{12}$ versus the neutrino
mass $m_1$ for different values of $D_{23}$.
    The deviation parameter $D_{23}$ is set to its best fit value $\sim 0.066$, $1\sigma$ limit $0.09$ and $0.11$
    for the dashed, solid, dotted lines
     respectively. (Right)  The deviation parameter $D_{23}$ versus $D_{12}$ for different values of $m_1$: $m_1=0, 0.01, 0.015 ~eV$ for the dashed, solid, dotted lines
     respectively. The horizontal dashed lines represent the best fit value and the upper $3\sigma$ limit of $D_{12}$.}
        \label{m1D12}
\end{figure}

From Eqs. (\ref{Delta1}) and (\ref{Delta2}) we can calculate the
deviation parameters from tri-bimaximal mixing (\ref{TBMdeviation})
as follows
\bea%
s_{13}&=& \frac{v^2}{\Lambda^2} \langle H \rangle^2\frac{i
\sqrt{\frac{3}{2}} f_4 \l_2^2 (u_2-u_3)}{(m_1-m_3) \left(f_5^2
   u_1^2-f_4^2 u_2 u_3\right)}, \nonumber\\
D_{23}&=& -\frac{v^2}{\Lambda^2} \langle H \rangle^2\frac{i \sqrt{3}
f_4 \l_2^2 (u_2-u_3)}{2(m_1-m_3) \left(f_5^2
   u_1^2-f_4^2 u_2 u_3\right)}\nonumber\\
D_{12}&=& -(\frac{v^2}{\Lambda^2} \langle H
\rangle^2)^2\frac{(u_2-u_3)^2 f_4^2 \l_2^4 (m_1+m_2-2 m_3)}{3
(m_1-m_2)
   (m_1-m_3)^2 \left(f_5^2 u_1^2-f_4^2 u_2
   (u_3)\right)^2}.
   \eea%
Here we set Dirac phase $\delta=0$. Thus, one can write the following
relations %
\bea%
|s_{13}|&=&\sqrt{2}~ D_{23}\nonumber\\
D_{12}&=& \frac{4(m_1+m_2-2m_3)}{9(m_1-m_2)}~D_{23}^2. %
\eea%
From the first relation, one finds that $s_{13}\sim 0.13$ (lower
$3\sigma$ experimental limit)  if $D_{23}\sim 0.09$, which
corresponds to $1\sigma$ limit of atmospheric neutrino mixing angle
\cite{Capozzi:2013csa}. In addition, if $D_{23}\sim 0.11$ ($2\sigma$
limit), one gets $s_{13}\sim 0.155$ (best fit value). In
Fig.~\ref{m1D12} we plot the relation between the lightest neutrino
mass $m_1$ and $D_{12}$ for $D_{23}$ is given by  its best value $0.066$, and $1\sigma$ limits $0.09$ and $0.11$. 
As can be seen from this figure, for $D_{23}=0.09-0.11$, which lead to
a consistent $s_{13}$,  the mass spectrum of neutrino should be
strongly hierarchical,  {\it i.e.} $m_1<0.01~eV$, in order to get
$D_{12}$ in the allowed range.  We also plot 
$D_{12}$ versus $D_{23}$ for different values of $m_1$, namely $m_1=0,~ 0.01~ eV,~ 0.015~eV$. It confirms the
same conclusion that the allowed range for $D_{12}$ can be achieved
if $m_1 \lsim 0.01 ~eV$ for $0.09 \lsim D_{23} \lsim 0.11$.

The approximated neutrino mass eigenvalues are given by%
\bea%
m_1 &\simeq & -3 \frac{v^2}{\Lambda^2} \langle H \rangle^2
\Big(\frac{2 \lambda_2^2}{\sqrt{4 f_5^2 u_1^2 + f_4^2 (u_2 -
u_3)^2}+f_4 (u_2 + u_3)}\Big),\nonumber\\
m_2 &\simeq &-3
\frac{v^2}{\Lambda^2} \langle H \rangle^2
\Big(\frac{\lambda_1^2}{f_1 u_1}\Big) ,\\
m_3 & \simeq & -3
\frac{v^2}{\Lambda^2} \langle H \rangle^2 \Big({\frac{2
\lambda_2^2}{\sqrt{4 f_5^2 u_1^2 + f_4^2 (u_2 -
u_3)^2}-f_4 (u_2 + u_3)}} \Big) \nonumber%
\eea%
We can tune the parameters to obtain the various mass hierarchy
spectra as the following: The normal hierarchy is achieved if $f_4\sim f_5\gg f_1$, $u_1\sim (u_2+ u_3)$ and the couplings $\lambda_s$ are of the same order. The degenerate scenario is obtained if $f_1\sim f_5\gg f_4$ and $u_1\gg u_2 , u_3$.
The inverted hierarchy is obtained if $f_4\sim f_5\gg f_1$ and $u_1\sim -(u_2+ u_3)$.

\begin{figure}[t]
    \psfrag{s13}{$s_{13}$}
    \psfrag{m}{$m_3$ (ev)}
    \includegraphics[height=6cm,width=8cm]{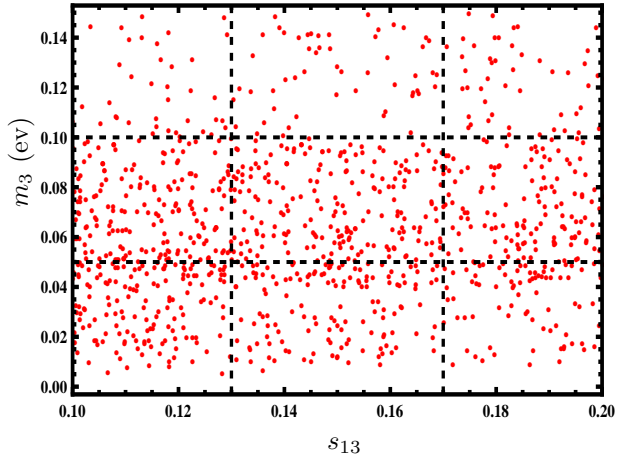}
    \caption{$\sin \theta_{13}$ versus the greatest neutrino mass $m_3$. All free parameters of the model are varying within their allowed regions}
    \label{m3s13}
\end{figure}

To ensure that there are various values of parameter space (consists of flavon
VEVs and couplings) that can account for the recent value of mixing
angle $\theta_{13}$ and neutrino masses simultaneously, we show in
Fig.\ref{m3s13} a correlation between $s_{13}$ and $m_3$, where all
free parameters vary in their allowed ranges.  In Fig.\ref{m3s13} we
display the deviation of solar neutrino parameters $D_{12}$ versus
the light neutrino mass $m_1$ and the deviation of atmospheric
neutrino parameter $D_{23}$.

It is important to notice the essential role of $S_2$ symmetry, which permutes the second flavour to the third one, that leads
to equal couplings in the Dirac mass matrix and right handed mass matrix. In this case, the neutrino mass matrix has the form of Eq.(\ref{right mass}) and hence tri-bimaximal mixing is realized. When we break the $S_2$ spontaneously by imposing different VEVs to the second and the third flavour of the flavon $\eta$, the deviation from tri-bimaximal is achieved.

\section {Quark masses and CKM mixing}

In this section we analyse the quark masses and mixing in the
framework of the symmetry group $\Delta(27)\ltimes S_2\times Z_4 $.
The quark transformations under $\Delta (27)$ are shown in Table
\ref{table3}. We also assume that the left handed quarks and up
right quarks transform under $S_2$ such that $Q_{L_2}
\leftrightarrow Q_{L_3}$ and $u_{R_2} \leftrightarrow u_{R_3}$ while
the down right quarks transform trivially under $S_2$ symmetry. From
these charge assignments, one finds that Yukawa interaction terms of
the up quarks, invariant under $\Delta (27)\ltimes S_2\times Z_4$,
are given by
\bea%
{\cal L}_u=\frac{1}{\Lambda}h^u_i\bar{Q}H u_R \eta_i,
\eea%
where $i=1,2,3$. Thus, the following  invariants terms are explicitly found: %
\bea%
&&\frac{1}{\Lambda}h^u_1 H(\bar{Q}_1 u_R+\bar{Q}_2 c_R+\bar{Q}_3 t_R)\eta_1,\nonumber\\ %
&&\frac{1}{\Lambda}h^u_2 H(\bar{Q}_1 u_R+ \omega^2\bar{Q}_2 c_R+\omega\bar{Q}_3 t_R)\eta_2,\nonumber\\
&&\frac{1}{\Lambda}h^u_3 H(\bar{Q}_1 u_R+ \omega\bar{Q}_2
c_R+\omega^2\bar{Q}_3 t_R)\eta_3. \eea %
\begin{table}[t] \begin{tabular}{|c|c|c|c|c|c|c|c|c|c|c|}
  \hline
  ~ Fields ~ & ~ Q ~ & ~ $d_R$ ~ & ~ $s_R$ ~ & ~ $b_R$ ~ & ~ $u_R$  \\
 \hline
 $\Delta (27)$&3 & $1_1$&$1_1$&$1_1$ & $3$    \\

  \hline $Z_4$ & 1 &1&-1&i &-1\\
  \hline
\end{tabular} \caption{Quark assignments under $\Delta (27)$ and
$Z_4$.} \label{table3}
\end{table} %
From the $S_2$ symmetry, $h^u_2=h^u_3$.
Therefore, the masses of the up quarks are given by:%
\bea%
m_u&=&\frac{1}{\Lambda}\langle H \rangle (h^u_1 u_1+h^u_2
(2u_2+\delta)),\nonumber\\m_c&=&\frac{1}{\Lambda}\langle H \rangle
(h^u_1 u_1+ h^u_2(\omega u_2+\omega^2
(u_2+\delta))),\nonumber\\m_t&=&\frac{1}{\Lambda}\langle H \rangle
(h^u_1
u_1+h^u_2(\omega^2  u_2+\omega ( u_2+\delta))),%
\eea%
where $\delta=u_3-u_2$. In general the coupling constants $h^u_i$
and VEVs $u_i$ are complex, so the previous three masses are
different and can account for the hierarchial mass spectrum of the
up quark sector. For instance, if $\frac{ u_i }{\Lambda}\sim {\cal
O}(\lambda_C^2)$ and $h^u_1\simeq ~ 6.85, ~h^u_2\simeq-6.85~ e^{i
\pi/3}, ~\frac{\delta}{\Lambda}\simeq -0.083 ~e^{i \pi/6}$, we can
get the up quark masses consistent with the following experimental
results: \be m_u(1 {\rm GeV}) & = & 4.5 \pm 1 ~{\rm MeV}, ~~~~~~
m_c(m_c) = 1.25 \pm 0.15 ~{\rm GeV}, ~~~~~~ m_t(m_t) = 166 \pm 5
~{\rm GeV}. \ee

Finally, we consider the down quark mass and mixing. From the charge assignments given in Table \ref{table3},  one can
write the following invariants:
\bea%
{\cal L}_d=\frac{1}{\Lambda^3}h_d\bar{Q} H d_R \phi^2 \sigma+\bar{Q}
H s_R (\frac{1}{\Lambda^2} h_{s_1}\xi^2+\frac{1}{\Lambda^3}
h_{s_2}\sigma^2 \eta_1)+\bar{Q} H b_R
(\frac{1}{\Lambda}h_{b_1}\phi+\frac{1}{\Lambda^3}h_{b_2}\chi^2 \xi).
\eea%
If $h_d\sim h_{s_1} \sim h_{b_1} \sim h_{b_1}\sim {\cal O}(1)$ while
$h_{s_2}\sim {\cal O}(0.1)$, then after spontaneous symmetry breaking
the following mass matrix of down quarks is obtained:

\bea%
m_d\simeq\left(
  \begin{array}{ccc}
    \lambda_C^4 & \lambda_C^3 & \lambda_C^4 \\
    \lambda_C^4 & \lambda_C^2 & \lambda_C^4 \\
    \lambda_C^4 & \lambda_C^3 & 1 \\
  \end{array}
\right)h_{b_1}\lambda_C^2\langle H\rangle,
\eea%
which can be diagonalized by two matrices,%
\bea%
\!\!m_d\!\!&\!\!=\!\!&\!\!V_L^T~ m_u^{diag} V_R\nonumber\\
\!\!\!&\!\!\!=\!\!&\!\!h_b\lambda^2_C \langle H\rangle\!\!\left(\!
     \begin{array}{ccc}
       1-\lambda_C^2 \!&\! \lambda_C \!&\! \lambda_C^4 \\
       -\lambda_C \!&\! 1-\lambda_C^2 \!&\! \lambda_C^4 \\
       -\lambda_C^4/2 \!&\! -\lambda_C^4 \!&\! 1 \\
     \end{array}
   \!\right)\!\left(\!
            \begin{array}{ccc}
              \lambda_C^4 \!&\! 0 \!&\! 0 \\
              0 \!&\! \lambda_C^2 \!&\! 0 \\
              0 \!&\! 0 \!&\! 1 \\
            \end{array}
          \!\right)\!\left(\!
           \begin{array}{ccc}
             1 \!&\! \lambda_C^2 \!&\! \lambda_C^4 \\
             -\lambda_C^4 \!&\! 1 \!&\! \lambda_C^6 \\
             -\lambda_C^4 \!&\! -2\lambda_C^6 \!&\! 1 \\
           \end{array}
         \!\right)\!.~~~~
         \eea

It is clear that the left handed rotation matrix $V_L$ is close to the quark mixing matrix,$V_{CKM}$, and the hierarchical spectrum of down quark
masses is obtained with the following mass ratios:
\bea%
m_b: m_s~ : ~m_d ~\approx~ 1~:~ \lambda_C^2 ~: ~\lambda_C^4,
\eea%
which are compatible with measured down quark masses:
\be
m_d(1 {\rm GeV}) & = & 8.0 \pm 2 ~{\rm MeV}, ~~~~~~ m_s(1 {\rm GeV}) =  150 \pm 50 ~{\rm MeV}, ~~~~~~  m_b(m_b) =  4.25 \pm 0.15 ~{\rm GeV}.
\ee

\section{Conclusions}
In this paper we have constructed a model of fermion masses and mixing based on an extension of the SM with a discrete flavour symmetry $\Delta(27)$. Our study is different from the previous $\Delta(27)$ analyses in two main points: $i)$ Our model is FCNC free, since one Higgs doublet is used to break the electroweak symmetry and SM singles only are involved in spontaneous breaking of $\Delta(27)$ $ii)$ Both quark and lepton masses and their mixing are simultaneously analysed under the same flavour symmetry. In fact most of the work in the literature focuses on the lepton sector only.

By assigning lepton doublets to $\Delta(27)$ triplet and right-handed leptons to singlets, we have shown that the charged lepton mass matrix is almost diagonal with the desired hierarchy.  Therefore, the MNS lepton mixing matrix is generated from the neutrino sector. We also argued that deviation from tri-bimaximal is due to spontaneous violation of the imposed $S_2$ symmetry. Similarly by assigning quark doublets and right-handed up quarks to $\Delta(27)$ triplets and right-handed down quarks to singlets, we obtained diagonal up-quark mass matrix and CKM quark mixing matrix arises from down sector only.

Finally, our model predicts that for $\sin \theta_{13} \simeq 0.13$, the mass of lightest neutrino is  $\lsim{\cal O}(0.1)$ eV and $\sin^2 \theta_{23} \simeq 0.41$, which is a remarkable deviation from maximal mixing.

\section*{Acknowledgements}
This work was partially supported by ICTP grant AC-80. S.K. would like to acknowledge partial support by
European Union FP7 ITN INVISIBLES (Marie Curie Actions, PITN-GA-2011-289442).
He would also like to thank Physics Department at Southampton University for hospitality where part of this work took place.


\end{document}